\documentclass{appolb}
\usepackage{graphicx}

\def\gsim{\mathrel{\rlap{\lower4pt\hbox{\hskip1pt$\sim$}}
 \raise1pt\hbox{$>$}}}

 \newcommand\la{\langle}
 \newcommand\ra{\rangle}
 \newcommand\beq{\begin{equation}}
 
 \newcommand\eeq{\end{equation}}
 \newcommand\beqn{\begin{eqnarray}}
 \newcommand\eeqn{\end{eqnarray}}

\def\GeV{\,\mbox{GeV}}

\def\lsim{\mathrel{\rlap{\lower4pt\hbox{\hskip1pt$\sim$}}
    \raise1pt\hbox{$<$}}}         
\def\gsim{\mathrel{\rlap{\lower4pt\hbox{\hskip1pt$\sim$}}
    \raise1pt\hbox{$>$}}}         

\def\GeV{\,\mbox{GeV}}

\begin{document}
\title{
Modeling photon radiation in soft hadronic collisions 
\thanks{Presented at ``Diffraction and Low-$x$ 2022'', Corigliano Calabro (Italy), September 24-30, 2022.}%
}
\author{B.~Z.~Kopeliovich,  I.~K.~Potashnikova
\address{Departamento de F\'{\i}sica,
Universidad T\'ecnica Federico Santa Mar\'{\i}a,
Avenida Espa\~na 1680, Valpara\'iso, Chile}
 \\[3mm]
{M.~Krelina
 \address{Czech Technical University in Prague, FNSPE, B\v rehov\'a 7, 11519
Prague, Czech Republic}
}
\\[3mm]
K.~Reygers
 \address{Physikalisches Institut, University of Heidelberg, Germany}
 }
\maketitle
\begin{abstract}
Soft hadronic collisions with multiple production of (anti)quarks accompanied with soft photon radiation are described in terms of higher Fock states of the colliding hadrons, which contain a photon component as well. The Fock state distribution functions are shaped with the Quark-Gluon String Model. Photon radiation by quarks is described within the color-dipole phenomenology. The results of calculations are in a good accord with available data in a wide range of transverse momenta of the photons.
\end{abstract}
  
It was demonstrated in \cite{kps-low} that the bremsstrahlung model (BM) \cite{BM}, used as a reference for comparison with the production rate of small-$k_T$ photons radiated in inelastic hadronic collisions at high energy, is incorrect, what led to so called soft photon puzzle (see e.g. in \cite{klaus}). 
  Therefore, an alternative description of soft photon radiation is required.
  
  \section{Parton model description at a hard scale}
  
Within the parton model radiation of a heavy photon of mass $M$ (Drell-Yan) in the target rest frame, based on the factorization theorem, has the form,
\beqn
\frac{d^4\sigma}{dM^2dx_F dk_T^2}=\frac{\alpha_{em}}{3\pi M^2}\frac{x_1}{x_1+x_2}
&&\int\limits_{x_1}^1\frac{d\alpha}{\alpha^2} \sum\limits_f Z_f^2
\left\{q_f\left(\frac{x_1}{\alpha}\right)
+\bar q_f\left(\frac{x_1}{\alpha}\right)\right\}\nonumber\\
\times\frac{d\sigma(q_fN\to\gamma^*X)}{d\ln{\alpha} d^2k_T},
\label{DY}
\eeqn
with the standart notations, $\alpha=p_+^\gamma/p_+^q$; $x_1x_2=M^2/s$; $x_1-x_2=x_F$.

\begin{figure}[!htbp]
\centerline{
\includegraphics[width=4cm]{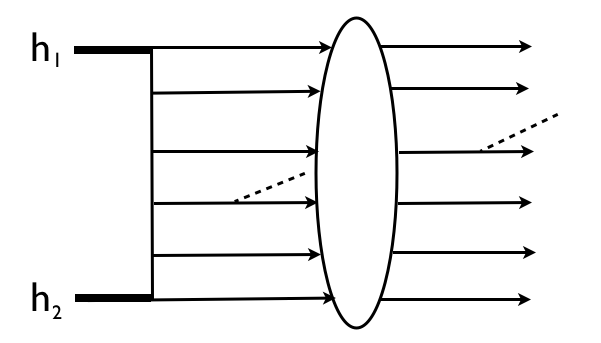}}
\caption{Space-time pattern of particle production at high energies.}
\label{fig:inelastic}
\end{figure}

The hard perturbative scale is imposed by the large invariant mass $M$ of the photon (dilepton). The sum of the (anti)quark distribution function in (\ref{DY}) is given by the well measured proton stricture function $F_2(x,M^2)$. The parton distribution functions in the colliding hadrons are illustrated by a parton comb in Fig.~\ref{fig:inelastic}.
  
\section{Parton model at a soft scale}  
  
Extrapolation of expression (\ref{DY}) to the soft regime is a challenge, since involves unknown nonperturbative effects. That can be done only within models. For the quark distribution function we rely on the popular and successful quark-gluon string model (QGSM) \cite{qgsm1,qgsm2}, or a similar dual parton model \cite{dual1,dual2}. Both models assume Regge behavior at the end-points $x\to 1$ or $x\to 0$, of the quark distribution functions, and a simple, but ad hoc, interpolation at medium $x$. We skip the simple, but lengthy expressions. The details can be found e.g. in \cite{qgsm2}.

The last factor in the radiation cross section (\ref{DY}) $d\sigma(q_fN\to\gamma X)/d\ln{\alpha}/ d^2k_T$ is calculated at the soft scale within the color dipole phenomenology 
\cite{BK1}-\cite{BK6}, adjusted to precise data on DIS from HERA,
\beqn
\frac{d\sigma(qN\to\gamma X)}{d\ln{\alpha} d^2k_T} &=&
\frac{1}{2\pi)^2}\int d^2r_1 d^2r_2\, \exp[i\vec k_T(\vec r_1-\vec r_2)]
\nonumber\\
&\times&
\Psi^*_{\gamma q}(\alpha,\vec r_1)\Psi_{\gamma q}(\alpha,\vec r_2)\,
\sigma_\gamma(\vec r_1,\vec r_2,\alpha),
\label{DM}
\eeqn
where
\beq
\sigma_\gamma(\vec r_1,\vec r_2,\alpha)=
{1\over2}\left\{\sigma_{\bar qq}(\alpha r_1)+
\sigma_{\bar qq}(\alpha r_2)-
\sigma_{\bar qq}[\alpha(\vec r_1-\vec r_2)]\right\}.
\label{sigma3}
\eeq
The quark-photon distribution function reads,
\beq
\Psi_{\gamma q}(\alpha,\vec r)=
\frac{\sqrt{\alpha_{em}}}{2\pi}\,
\chi_f \hat O\chi_i
K_0(\alpha m_q r), 
\label{psi1}
\eeq
and
\beq
\hat O = \vec{e^*}\left\{ im_q\alpha^2\left[\vec n\times\vec\sigma\right]+
\alpha\left[\vec\sigma\times\vec\nabla\right]-i(2-\alpha)\vec\nabla
\right\}.
\label{psi2}
\eeq

The $\bar qq$ dipole-nucleon cross section $\sigma_{\bar qq}(r)$ in (\ref{sigma3}) has been parametrized and  fitted to DIS and photoproduction data from NMC and HERA. The details can be found in \cite{BK3}.

Combining the QGSM distribution functions with the cross section (\ref{DM}) results in the radiation cross section, which is parameter free (we do not fit the data to be explained), either in the shape of $k_T$ distribution, or in the absolute values. We assumed a primordial transverse momentum distribution of the incoming quarks to have a Gaussian shape with $\sqrt{\la q_T^2\ra}=0.35\GeV$. Correspondingly the radiated photon acquires additional transverse momentum $\vec k^\prime_T=\alpha\vec q_T$. 

\section{Comparison with data}

\begin{figure}[!t]
\centerline{
\includegraphics[width=9cm]{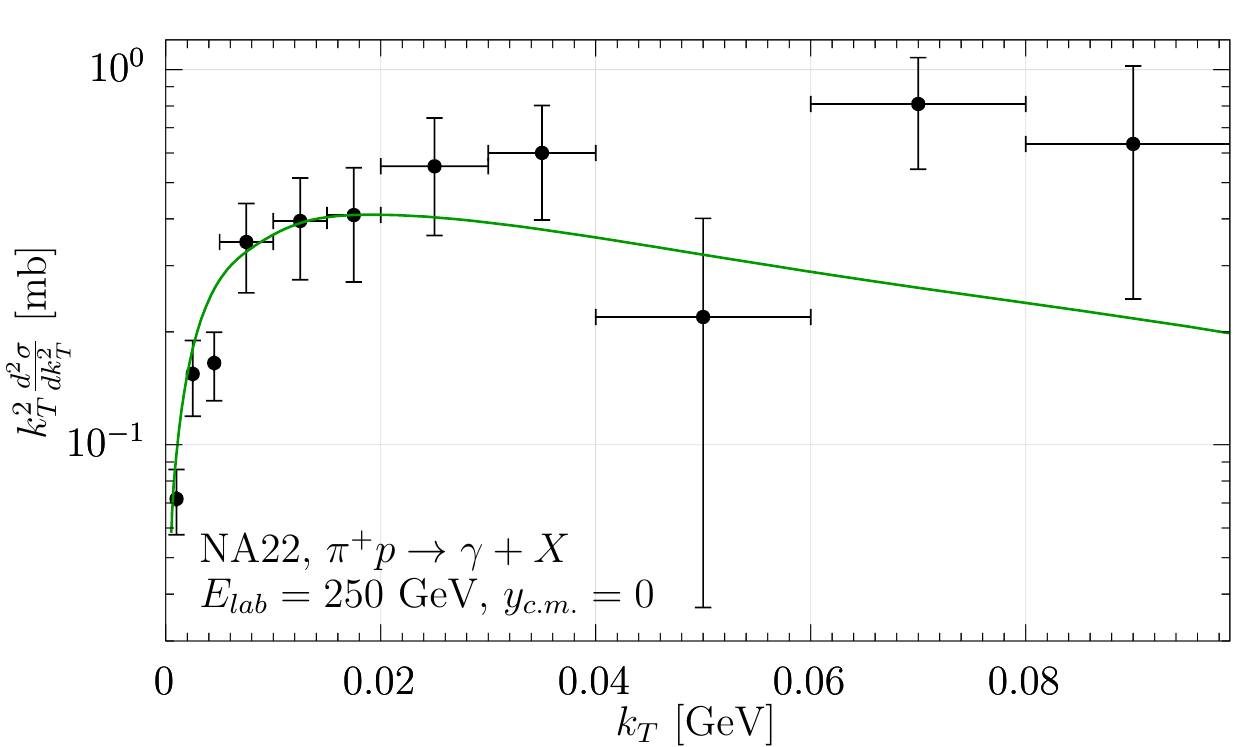}}
\caption{Comparison with data of the NA22 experiment \cite{NA22} for $\pi^+p\to\gamma X$ at  $E_{lab}=250\GeV$.}
\label{fig:fig1}
\end{figure}

\begin{figure}[hb]
\centerline{
\includegraphics[width=9cm]{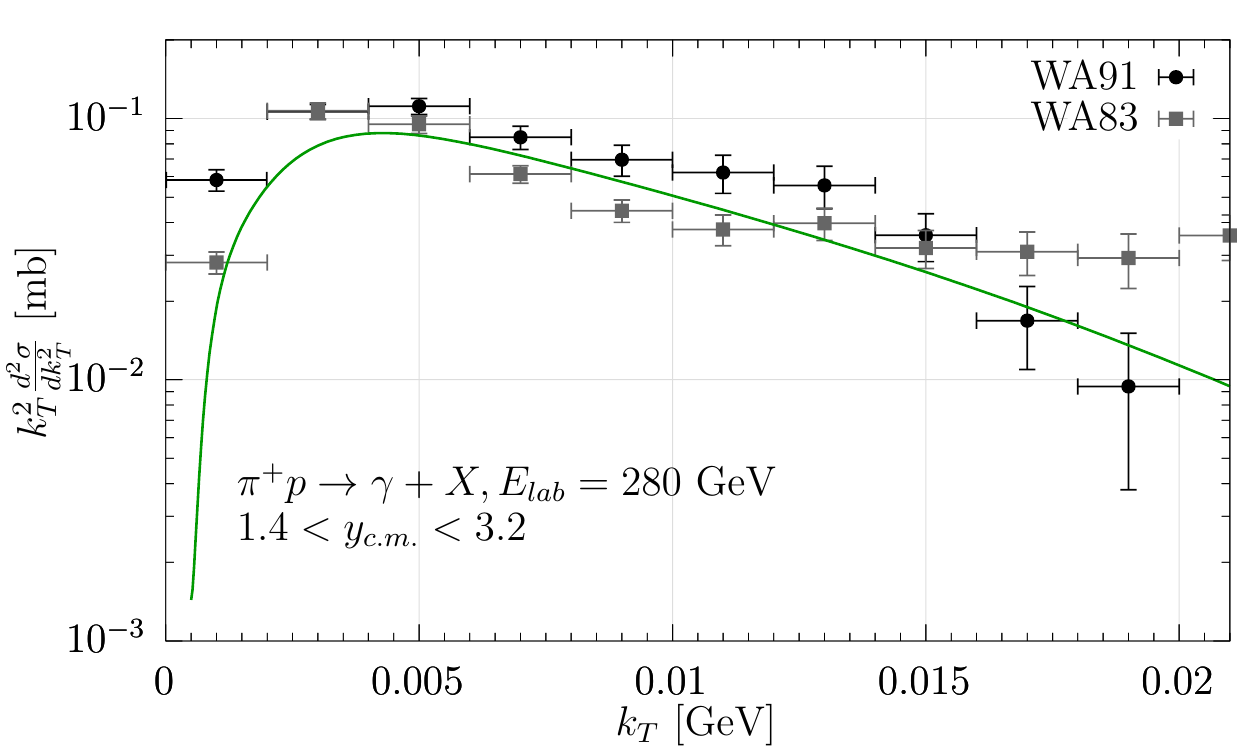}}
\caption{Comparison with data of the WA83 \cite{WA83} and WA91 \cite{WA91} experiments for $\pi^+p\to\gamma X$ at  $E_{lab}=280\GeV$.}
\label{fig:fig2}
\end{figure}

The results of calculations are compared with data on the radiative cross section of $\pi^+ p\to\gamma+X$ from the NA22 experiment at $E_{lab}=250\GeV$ in Fig.~\ref{fig:fig1}, and from WA91/WA83 experiments at $E_{lab}=280\GeV$ in Fig.~\ref{fig:fig2}.

We see no sizable deviation from data at small $k_T$, i.e. no anomalous enhancement of soft photons.

\begin{figure}[t]
\centerline{
\includegraphics[width=9cm]{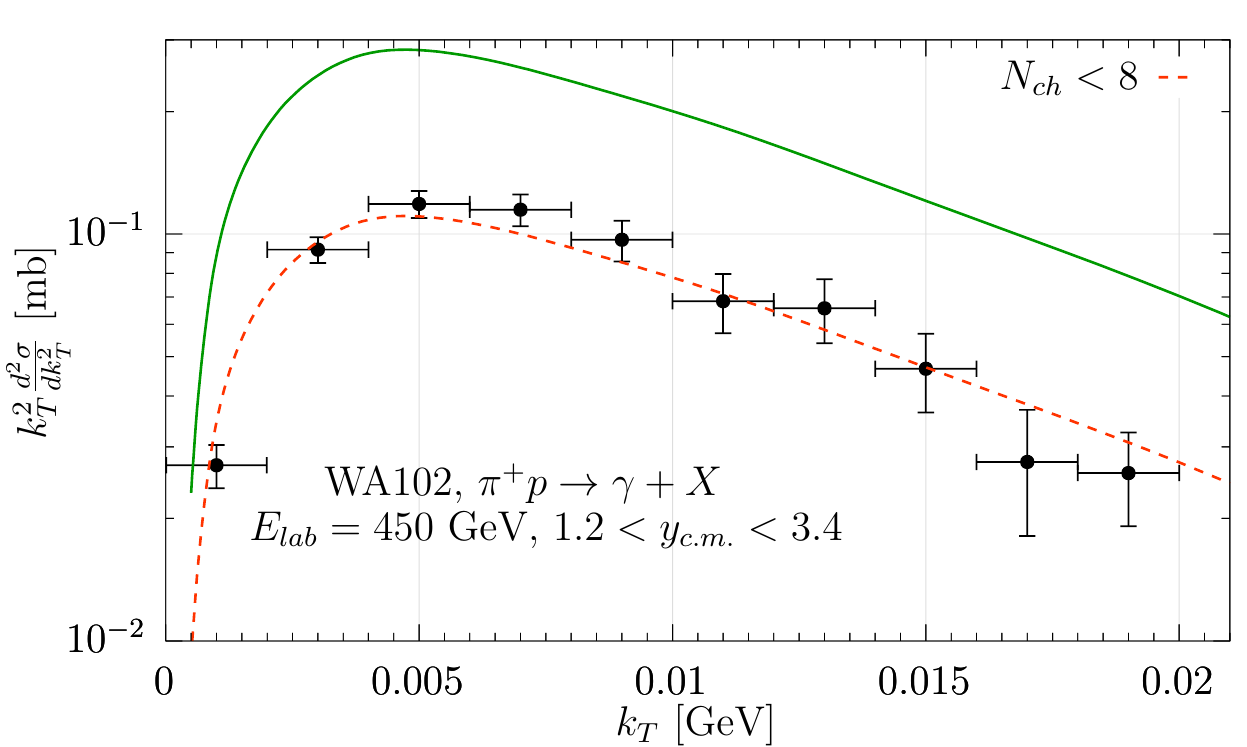}}
\caption{Comparison with data of the WA102 \cite{WA102} experiment for $\pi^+Be\to\gamma X$ at  $E_{lab}=450\GeV$.}
\label{fig:fig3}
\end{figure}

At somewhat higher energy $E_{lab}=450\GeV$ \cite{WA102} our calculations depicted by solid curve in Fig.~\ref{fig:fig3}, apparently overestimate the data of the WA102 experiment. 
However, the experiment had specific cuts, namely, events with number of charge tracks $N_{ch}>8$, were excluded. To calculate the multiplicity distribution we assume the Poisson distribution of the number of unitary cut Pomerons, and employed the result of QGSM \cite{kaid-ter}. So we obtained a suppression factor 
\beq
\delta=\frac{\sum\limits_{N_{ch}=0}^8}{\sum\limits_{N_{ch}=0}^\infty}=
0.39.
\label{Nch}
\eeq
Dashed curve, which incorporate this factor well agrees with data.

\section{Conclusions}
\begin{itemize}
\item
The observed enhancement of low-$k_T$ photons in comparison with incorrect calculations, should not be treated as a puzzle.
\item
The parton model description of photon radiation is extrapolated to to the soft scale regime.
The (anti)quark distribution functions are evaluated within the popular quark-gluon string model, based of Regge phenomenology.
\item
Soft photon bremsstrahlung by projectile quarks is calculated within the color-dipole model.
The quark-antiquark dipole cross section is fitted to DIS and soft photo-production data in a wide range of transverse dipole separations and energies.

\end{itemize}
{\bf Acknowledgements:}
This work of B.Z.K. and I.K.P. was supported in part by grant (Chile) ANID PIA/APOYO AFB220004.\\
The work of M.K. was supported by the project of the International Mobility of Researchers - MSCA IF IV at CTU in Prague 
CZ.02.2.69/0.0/0.0/20\_079/0017983, Czech Republic.

{}

\end{document}